\documentclass[twocolumn,twocolappendix]{aastex63}

\usepackage{multirow}
\usepackage{booktabs}

\usepackage{graphicx}
\definecolor{nblue}{HTML}{3465A4}
\definecolor{nred}{HTML}{EF4A53}

\definecolor{simone}{HTML}{619530}

\definecolor{tassos}{HTML}{5C3566}

\definecolor{nbrown}{HTML}{B56F2A}
\definecolor{ntour}{HTML}{249494}
\definecolor{ngray}{HTML}{888A85}
\definecolor{nlila}{HTML}{B13C6A}

\def\p{\varphi}
\def\d{\delta}

\def\Msun{\mathrm{M}_{\odot}}

\def\posydon{\texttt{POSYDON}}
\def\mesa{\texttt{MESA}}
\def\cosmic{\texttt{COSMIC}}

\def\Mhec{M_\mathrm{He,\,core}}
\received{...}
\revised{...}
\accepted{...}
\submitjournal{AJ}

\shorttitle{The observability of BHNS mergers as multi-messenger sources}
\shortauthors{Rom\'an-Garza et al.}
\graphicspath{{./}{figures/}}
\begin{document}

\title{The role of core-collapse physics in the observability of black-hole neutron-star mergers as multi-messenger sources}

\author[0000-0002-5962-4796]{Jaime\,Rom\'an-Garza}\thanks{jaime.roman@etu.unige.ch}
\affiliation{Observatory of Geneva, University of Geneva,
Chemin Pegasi 51, 1290,
Versoix, Switzerland.}

\author[0000-0002-3439-0321]{Simone\,S.\,Bavera}
\affiliation{Observatory of Geneva, University of Geneva,
Chemin Pegasi 51, 1290,
Versoix, Switzerland.}

\author[0000-0003-1474-1523]{Tassos\,Fragos}
\affiliation{Observatory of Geneva, University of Geneva,
Chemin Pegasi 51, 1290,
Versoix, Switzerland.}

\author[0000-0002-7464-498X]{Emmanouil\,Zapartas}
\affiliation{Observatory of Geneva, University of Geneva,
Chemin Pegasi 51, 1290,
Versoix, Switzerland.}

\author[0000-0003-4260-960X]{Devina\,Misra}
\affiliation{Observatory of Geneva, University of Geneva,
Chemin Pegasi 51, 1290,
Versoix, Switzerland.}

\author[0000-0001-5261-3923]{Jeff\,Andrews}
\affiliation{Center for Interdisciplinary Exploration and Research in Astrophysics (CIERA) and Department of Physics and Astronomy, Northwestern University, 1800 Sherman Avenue, Evanston, IL 60201, USA}

\author[0000-0002-0403-4211]{Scotty\,Coughlin}
\affiliation{Center for Interdisciplinary Exploration and Research in Astrophysics (CIERA) and Department of Physics and Astronomy, Northwestern University, 1800 Sherman Avenue, Evanston, IL 60201, USA}

\author[0000-0002-4442-5700]{Aaron\,Dotter}
\affiliation{Center for Interdisciplinary Exploration and Research in Astrophysics (CIERA) and Department of Physics and Astronomy, Northwestern University, 1800 Sherman Avenue, Evanston, IL 60201, USA}

\author[0000-0003-3684-964X]{Konstantinos\,Kovlakas}
\affiliation{Physics Department, University of Crete,
GR 71003, Heraklion, Greece}
\affiliation{Institute of Astrophysics, Foundation for Research and Technology-Hellas,
GR 71110 Heraklion, Greece}

\author{Juan\,Gabriel\,Serra}
\affiliation{Center for Interdisciplinary Exploration and Research in Astrophysics (CIERA) and Department of Physics and Astronomy, Northwestern University, 1800 Sherman Avenue, Evanston, IL 60201, USA}

\author[0000-0002-2956-8367]{Ying\,Qin}
\affiliation{Center for Interdisciplinary Exploration and Research in Astrophysics (CIERA) and Department of Physics and Astronomy, Northwestern University, 1800 Sherman Avenue, Evanston, IL 60201, USA}
\affiliation{Department of Physics, Anhui Normal University, Wuhu, Anhui 241000, China}

\author[0000-0003-4474-6528]{Kyle\,A.\,Rocha}
\affiliation{Center for Interdisciplinary Exploration and Research in Astrophysics (CIERA) and Department of Physics and Astronomy, Northwestern University, 1800 Sherman Avenue, Evanston, IL 60201, USA}

\author{Nam\,Hai\,Tran}
\affiliation{DARK, Niels Bohr Institute, University of Copenhagen,
Jagtvej 128, 2200 Copenhagen, Denmark}



\begin{abstract}

Recent detailed 1D core-collapse simulations have brought new insights on the final fate of massive stars, which are in contrast to commonly used parametric prescriptions. In this work, we explore the implications of these results to the formation of coalescing  black-hole (BH) -- neutron-star (NS) binaries, such as the candidate event GW190426\_152155 reported in GWTC-2. Furthermore, we investigate the effects of natal kicks and the NS's radius on the synthesis of such systems and potential electromagnetic counterparts linked to them. Synthetic models based on detailed core-collapse simulations result in an increased merger detection rate of BH-NS systems ($\sim 2.3\,\mathrm{yr}^{-1}$), 5 to 10 times larger than the predictions of ``standard'' parametric  prescriptions. This is primarily due to the formation of low-mass BH via direct collapse, and hence no natal kicks, favored by the detailed simulations. The fraction of observed systems that will produce an electromagnetic counterpart, with the detailed supernova engine, ranges from $2$--$25$\%, depending on uncertainties in the NS equation of state. Notably, in most merging systems with electromagnetic counterparts, the NS is the first-born compact object, as long as the NS's radius is $\lesssim 12\,\mathrm{km}$. Furthermore, core-collapse models that predict the formation of low-mass BHs with negligible natal kicks, such as the ones from detailed core-collapse studies, increase the detection rate of GW190426\_152155-like events to $\sim 0.6 \, \mathrm{yr}^{-1}$; with an associated probability of electromagnetic counterpart $\leq 10$\% for all supernova engines. However, increasing the production of direct-collapse low-mass BHs also increases the synthesis of binary BHs, over-predicting their measured local merger density rate. In all cases, models based on detailed core-collapse simulation predict a ratio of BH-NSs to binary BHs merger rate density that is at least twice as high as other prescriptions.

\end{abstract}


\section{Introduction} \label{sec:intro}

The recently released catalogue of the LIGO Scientific and Virgo Collaboration (LVC), GWTC-2, includes for the first time an event, GW190426\_152155, classified as a black hole (BH) -- neutron star (NS) merger \citep{Abbott:2020niy}. In addition, GW190814, an extreme mass-ratio merger event, has an estimated mass of 2.59$^{+0.08}_{-0.09}$ for the lower-mass compact object, making it unclear whether it is a binary BH (BBH) or a BH--NS (BHNS) merger \citep{2020ApJ...896L..44A}. Due to the relatively low significance of GW190426\_152155 and the unclear nature of GW190814, no BHNS merger rate density was estimated based on GWTC-2 \citep{2020arXiv201014533T}, with the older estimates from GWTC-1 setting only an upper limit of $<610\,\rm Gpc^{3}\,yr^{-1}$ \citep{2019PhRvX...9c1040A}. Nevertheless, taking into account that the first half of the LVC's third observing run (O3a) included 177.3 days of data suitable for coincident analysis and assuming 1 or 2 detections, one can estimate a detection rate of $\sim$2--4$\,\mathrm{yr}^{-1}$.

The first detection of a binary NS merger was accompanied by an electromagnetic counterpart (EMC),  which was observed in the whole electromagnetic spectrum as a kilonova and a short Gamma-ray burst \citep{2017ApJ...848L..12A, 2017ApJ...848L..13A}. The merger of a BH with a NS is also expected to be accompanied by a similar EMC, if the tidal disruption radius of the NS is outside the innermost stable circular orbit (ISCO) of the BH. The maximum mass of a non-spinning BH for this to happen, assuming a 1.4\,M$_{\odot}$ NS, is $\sim$3.5\,M$_\odot$, with the exact value depending on the adopted NS equation of state \citep[e.g.][]{capano2020stringent}. If, however, the BH is spinning, then the ISCO moves closer to the BH and the corresponding BH mass limit becomes as high as $\sim 20\,\rm M_{\odot}$ for a maximally spinning BH \citep[see][]{foucart2018remnant}, vastly increasing the probability of an electromagnetic counterpart.

Under the assumption of efficient angular momentum transport in the interior of stars, the first-born compact object in a binary is expected to be formed with negligible spin \citep{2015ApJ...800...17F,qin2018spin}. This is because as the progenitor star expands to become a supergiant, most of its angular momentum is transported to its outer layers, which are then removed via winds and Roche-lobe overflow. On the other hand, the immediate progenitor of the second-born compact object, in the isolated binary formation channels, is a stripped helium (He) star in a close orbit with its first-born compact-object companion. There, the He-star has a chance to be spun up via tides, and thus give rise to a compact object with a significant spin \citep[e.g.,][]{van2007long,qin2018spin,2020A&A...635A..97B,2020arXiv201016333B}. Hence, the only way to have a highly spinning BH in a BHNS system is to have the NS be the first-born compact object, which will then tidally spin up the BH's progenitor star.

%\textbf{**Mention that there was no electromagnetic counterpart observed, , mention also \citep{2020arXiv201014550T}, }

Coalescing BHNSs formed via isolated binary evolution are thought to be sufficiently abundant, with theoretical estimates of their merger rate density covering the whole range of 0.1--1000\,Gpc$^3$\,yr$^{-1}$ \citep{giacobbo2018progenitors,2020A&A...636A.104B,drozda2020black}. Among them, however, BHNSs {in which} the NS is the first-born system should be rare for two main reasons: (i) the initially more massive primary star is typically the progenitor of the BH which tends to form first, and (ii) even if binary interactions reverse the pre-core-collapse mass of the primary and secondary, kicks imparted on newly-born NSs \citep{hobbs2005statistical}, which are typically higher than those on BHs, tend to disrupt the binary when the system is in a wide orbit, as is typically the case when the first compact object is formed.

The details of the general statements above depend crucially on the physics of core-collapse and compact object formation. The vast majority of binary population synthesis studies use the ``rapid'' and ``delayed'' mechanisms \citep{fryer2012compact} to prescribe the fate of massive stars. Both of them are parametric descriptions of the convection-enhanced supernova (SN) engine driven by neutrino losses \citep[see][]{herant1994inside}. In contrast to the ``dealayed'', the ``rapid'' prescription predicts a mass gap between BHs and NSs due to stronger convection which allows instabilities to grow rapidly and produces more energetic SN explosions.  Furthermore, both mechanisms predict a unique boundary in the core mass of the pre-core-collapse star that leads to the formation of a BH or a NS, above or below that boundary respectively.

Detailed 1D core-collapse simulations, on the other hand, show that there is no unique boundary in the core mass of the pre-core-collapse star that transitions between the formation of NSs and BHs. Instead, these studies find successive islands of successful and failed explosions leading to the formation  of NSs and BHs via direct collapse \citep[e.g.,][]{pejcha2015landscape,sukhbold2016core,ertl2020explosion,patton2020towards,Schneider:2020vvh,2012ASPC..453...91U,o2011black}.  This is the result of the non-monotonic behaviour between the central carbon-burning phase and the final core properties, linked to the convective episodes developed during the burning phase \citep[e.g.][]{2018ApJ...860...93S}. Another significant difference between these more recent calculations and the ``rapid'' and ``delayed'', is in the formation of BHs via successful explosions and significant fall-back. Whereas in the ``rapid'' and ``delayed'' prescriptions there is a wide range of pre-core-collapse core masses that lead to the formation of BHs via accretion of fall-back mass from a successful explosion. Detailed 1D simulations find that these cases are very rare and virtually all BHs are formed via direct collapse. The latter becomes important in the context of population synthesis studies, where the natal kicks imparted on BHs are most often normalized to the fraction of fall-back mass, while BHs formed via direct collapse receive no kick \citep[e.g.,][]{2008ApJS..174..223B}.

In this work we explore the effects of the core-collapse mechanism on the formation of coalescing BHNSs, such as GW190426\_152155, and their potential observability as electromagnetic transients. For the first time we consider a core-collapse prescription based on detailed 1D core-collapse simulations and study the effect of non-monotonic stellar explodability, with respect to the pre-collapse mass of the core. Finally, we discuss what we can learn in the future in terms of formation pathways, core-collapse physics and NS equation of state, once observations put a firmer constraints on the BHNS merger rate density and the fraction of them accompanied by an EMC.

\section{Methods \label{sec:methods}}

We use the software framework \posydon{} (Fragos et al. 2021, in prep.) to evolve populations of binaries for this study. \posydon{} allows, among other functionalities, to couple parametric population synthesis codes with such models for different phases of a binary's evolution. In this work, we coupled the parametric code \texttt{COSMIC} \citep{breivik2019cosmic} to evolve binaries from the zero-age main sequence (ZAMS) until the first compact object strips its companion; and a grid of $\sim$170,000 detailed binary evolution models ran with the \texttt{MESA} code \citep{2011ApJS..192....3P,2013ApJS..208....4P,2015ApJS..220...15P,2018ApJS..234...34P,2019ApJS..243...10P}  to follow the final evolutionary phase of a BHNS progenitor, i.e. that of a binary consisting of stripped He-star and a compact object in a close orbit. This allows us to accurately predict the spin of the second born compact object  \citep[see][for a detailed description of the simulation setup]{2020arXiv201016333B}.

%\subsection{Population synthesis}

For all populations evolved, we consider the following initial binary properties: the mass of the most massive star $m_1$ is distributed by the \cite{kroupa2001variation} initial mass function in a mass range of $[5,150]\,\Msun$. We assume the mass ratio at birth is distributed uniformly as $q\in[0,1]$ \citep[][]{sana2012binary}. The initial orbital periods are distributed in the range $[0.4, 10^{5.5}]$ days as in \cite{sana2012binary}, extending the distribution for for low values with a flat distribution as in \cite{2020arXiv201016333B}. All binaries have zero birth eccentricity, and we assume an overall binary fraction $f_b=0.7$ \citep[see][]{2020arXiv201016333B,sana2012binary}. For each model, we evolve $5\times10^6$ binaries per metallicity, for 10 different metallicity values, $Z\in\ $[0.0001, 0.0002, 0.0003, 0.0006, 0.001 , 0.0018, 0.0031, 0.0055, 0.0098, 0.0174]. This corresponds to a total stellar mass of $\sim 5\times 10^8 \Msun$ for the underlying stellar population, per metallicity bin; and corresponds to a fraction of the initial mass function of $f_\mathrm{corr} = 0.212$.

We use the same set of physical model parameters as in \cite{2020arXiv201016333B}. Specifically, we adopt mass-transfer (MT) stability according to the values of $q_\mathrm{crit}$ as described there. Unstable MT is modelled with the classical $\alpha_\mathrm{CE}-\lambda$ common-envelope (CE) formalism \citep{1976IAUS...73...35V, 1984ApJ...277..355W}, where assume an $\alpha_\mathrm{CE} = 1$ and $\lambda_\mathrm{CE}$ fits from \cite{claeys2014theoretical} without taking into account the ionization energy of the envelope. Moreover, we assume the pessimistic common-envelope scenario, namely all systems that start common envelope evolution with a star in the Hertzsprung's gap are considered to merge due to the unsuccessful envelope ejection \citep{2004ApJ...601.1058I,2007ApJ...662..504B}. The electron-capture SN (ECSN) prescription described in \cite{podsiadlowski2004effects} is used, which maps Helium-core masses  in the range $[1.4,2.5]\,\Msun$ to remnant baryonic mass $1.38\, \Msun$ as in  \cite{giacobbo2020revising}. For the pair-instability and pulsational pair-instability SNe we consider the prescription by \cite{marchant2019pulsational} which limits the maximum BH mass at $\sim 44 \, \Msun$. 

To model the core-collapse, in addition to the ``rapid'' and ``delayed'' mechanisms by \citet{fryer2012compact}, we implement a new prescription based on detailed 1D core-collapse simulations. We use the publicly available data on pre-SN models and the remnant properties produced by the N20 engine of \cite{sukhbold2016core}. We consider the He-core mass of the star at the pre-SN phase, $\Mhec$, to predict the remnant's baryonic mass, $M_\mathrm{rem,\,bar}$, taking into account whether the SN explosion is predicted to be successful or not. We only consider the successful explosions that will produce NSs, as the ones that produce BHs are rare. A complete description of the implementation of the engine in our population synthesis study is in the Appendix~\ref{app:N20}. 

For each mechanism three populations were produced with the assumptions described in Table~\ref{table:populations}. Our fiducial model, called STANDARD, considers a Maxwellian distribution  with $\sigma_\mathrm{ECSN}=20$ km/s for kicks imparted on NS formed from ECSN and $\sigma_\mathrm{FeCCSN}=265$ km/s for Fe core-collapse SN (FeCCSN); the kicks in this model are fallback-weighted as in \cite{fryer2012compact}. The FULL-ECSN-KICK model considers that NS formed from ECSN and FeCCSN receive the same kicks ($\sigma_\mathrm{ECSN}=\sigma_\mathrm{FeCCSN}=265$ km/s). Finally, the NO-BH-KICK model considers BHs receive no natal kicks, even if they are not produced by direct collapse.

While the  evolution of binaries from  ZAMS to the formation of a compact object plus a He-star was computed with the parametric code \cosmic, the last phase of the evolution of a close compact object plus He-star was performed by interpolating a grid constituted by 172,570 detailed \texttt{MESA} binary evolution models. This last step, allows us to derive accurate predictions of the second-born compact object's spin \citep{qin2018spin,2020A&A...635A..97B}. The details on the grid and its interpolation are discussed in Appendix~\ref{app:grid}.

We extract the BHNS mergers,  systems that merge due to gravitational wave radiation emission in less time than the current age of the Universe, and compute the number of BHNS mergers per unit mass. The merging timescale, by gravitational wave radiation, is computed according to \cite{peters1964gravitational}.  Adopting the $\Lambda$CDM cosmology, we distribute the BHNS across the metallicity-dependent cosmic star-formation history; assuming metallicities follow a truncated log-normal distribution with standard deviation $0.5~\mathrm{dex}$ around the empirical mean metallicity function derived by \citet{madau2017radiation}. We compute the merger rate densities and detection rates as in \cite{2020arXiv201016333B}, assuming the simulated ``mid high/late low'' LVC O3  detector sensitivity \citep{abbott2018prospects}, by considering a single detector signal-to-noise ratio threshold $> 8$ that simulates a 2 network detector \citep[see][]{barrett2018accuracy}. Finally, we compute the fraction of events that produce EMCs. We assume that a BHNS merger will produce an electromagnetic counterpart if the mass outside the BH ISCO after the merger, $M^\mathrm{ejecta}$, is greater than zero. The detailed description on the computation of $M^\mathrm{ejecta}$ is described in Appendix~\ref{app:EMC}. Taking into account uncertainties in NS equation of state \citep[e.g.][and references there in]{capano2020stringent,chatziioannou2020neutron}, we consider three different constant values for the NS readius, $R_\mathrm{NS} \in [11, 12, 13]$\,km, to compute $M^\mathrm{ejecta}$ and, hence, predict the occurrence EMCs.

%%%%%%%%%%%%%%%%%%%%%%%%%%%%%%%%%%%%%%%%%%%%%%%%% 

\begin{table}
\caption{Definition of the models used to evolve the binary stellar populations. \label{table:populations}}
\centering
\begin{tabular}{ c c c c c}
%\toprule

  Model    &  $\sigma_\mathrm{ECSN}$&   $\sigma_\mathrm{FeCCSN}$ &  NS  &  BH  \\
    name    & [km/s]& [km/s]  &  kicks &  kicks \\

\toprule

 STANDARD& 20 & 265 &    fallback &    fallback \\
  &  &  &    weighted &    weighted \\

\midrule

 FULL-ECSN& 265 & 265 &    fallback &    fallback \\
  -KICK&  &  &    weighted &    weighted \\

\midrule

NO-BH& 20 & 265 &    fallback &    No \\
  -KICK &  &  &    weighted &    kicks \\

\bottomrule
\end{tabular}
\end{table}

%%%%%%%%%%%%%%%%%%%%%%%%%%%%%%%%%%

%%%%%%%%%%%%%%%%%%%%%%%%%%%%%%%%%%%%%%%%%%%%%%
\begin{figure*}
\centering
%\captionsetup{justification=justified}

\includegraphics[width=0.49\textwidth]{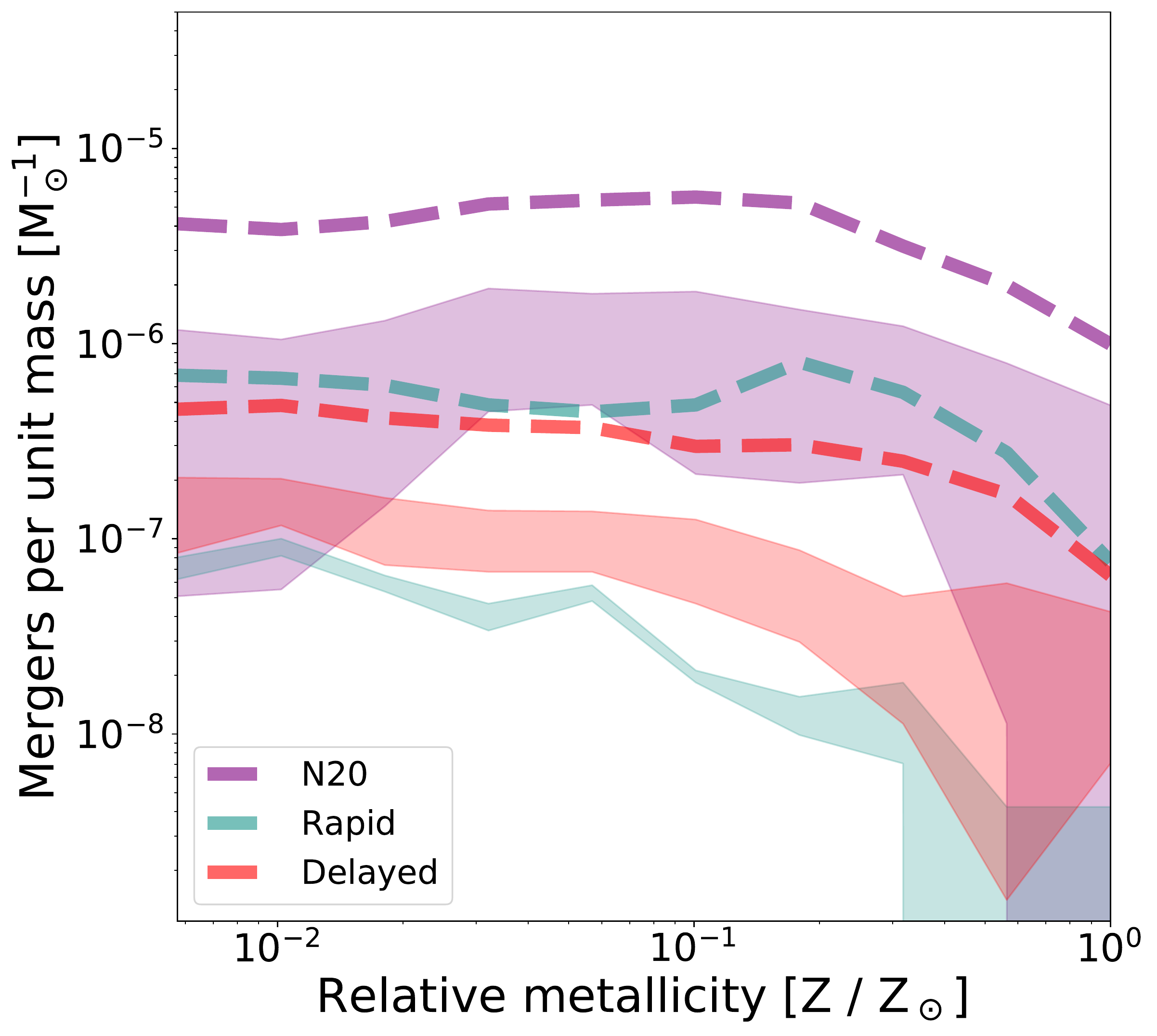}
\includegraphics[width=0.49\textwidth]{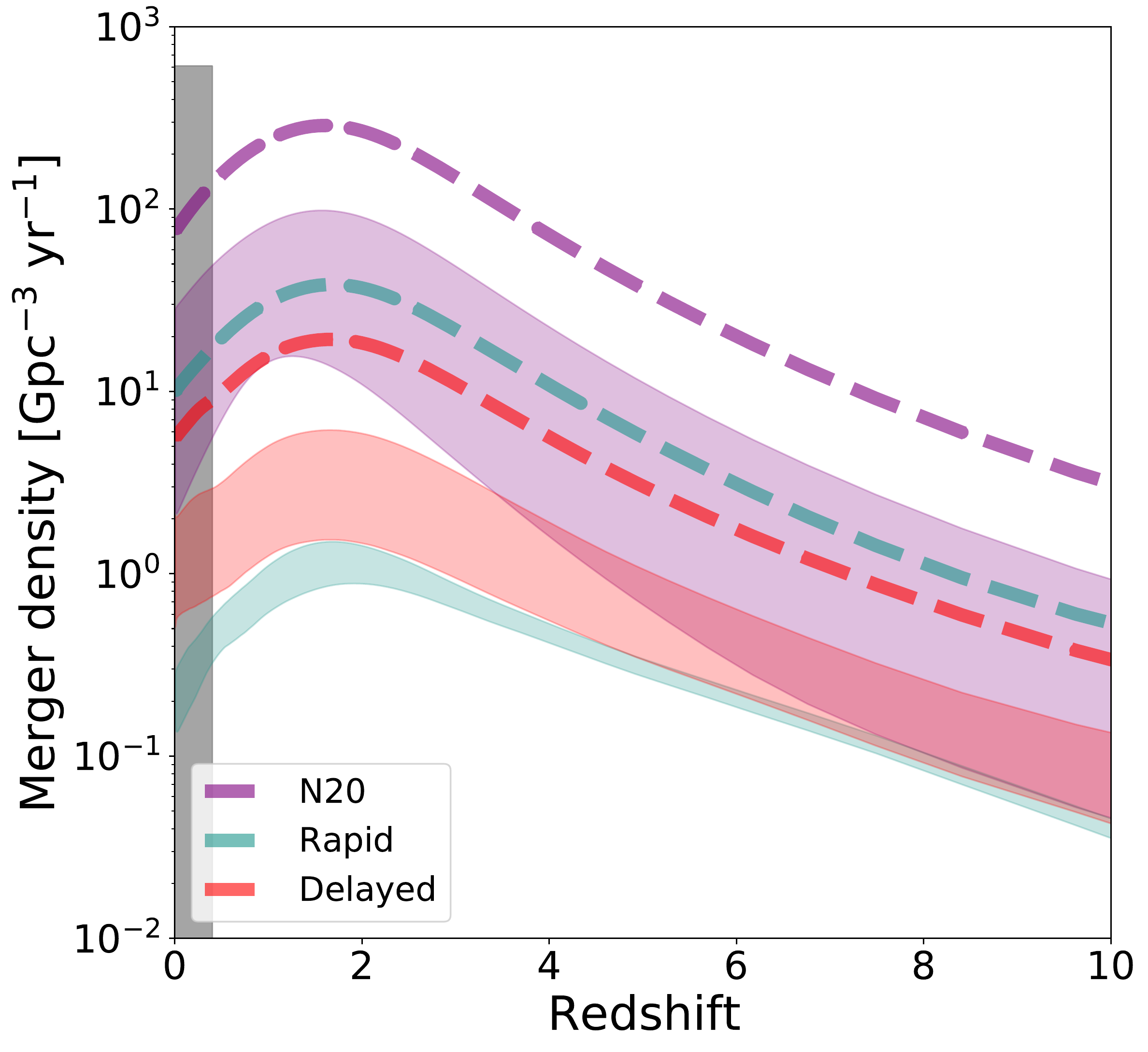}

\caption{BHNS mergers per unit mass in terms of relative metallicity to the solar value (left) and its translation to the merger density rate history as a function of redshift $z$ (right) for the STANDARD model. The dashed lines represent the whole population of BHNSs mergers, while the contours enclose the number of electromagnetic counterparts by considering the three values of NS radius as described in Sec.~\ref{sec:methods}; where upper and lower boundaries of the contours are the predictions taking into account $R_\mathrm{NS} = 13$km and $R_\mathrm{NS} = 11$km respectively. In the right panel, the gray vertical bar is the $90\%$ confidence interval of local BHNS merger density rate reported in GWTC-1. }

 \label{fig:mergerhistory}
\end{figure*}
%%%%%%%%%%%%%%%%%%%%%%%%%%%%%%%%%%%%%%%%%%%%%%%%% 

\section{Results \label{sec:results}}

\subsection{The theoretical merger rates of BHNS}
The left panel of Fig.~\ref{fig:mergerhistory} shows the number of BHNS mergers per unit mass, namely the total evolved mass corrected by the initial mass function and the binary fraction, in terms of metallicity for the STANDARD model. In that panel the dashed lines represent the whole population of BHNSs for each core-collapse mechanism. While the shaded regions represent the subset of systems that will reproduce an EMC. The upper boundary of a shaded region is defined for $R_\mathrm{NS}=13$ km, while the lower boundary is delimited by considering $R_\mathrm{NS}=11$ km. By distributing the mergers on the cosmic star-formation history we find the merger density history in terms of redshift, shown in the right panel of Fig.~\ref{fig:mergerhistory}. There, we see that the local merger density for all core-collapse mechanisms is consistent with the upper limit reported on GWTC-1 \citep[see][]{2019PhRvX...9c1040A} and plotted as the gray shaded region. Furthermore, these results are consistent with other recent population synthesis studies \citep[e.g.][]{giacobbo2018progenitors, 2019MNRAS.490.3740N, drozda2020black, 2020A&A...636A.104B}.

We show the detection rate of BHNSs for simulated O3 LIGO/Virgo sensitivity for all the populations in the upper panels of Fig.~\ref{fig:rates}. The estimated BHNSs detection rates are plotted with unfilled diamonds, while the filled symbols represent the subset of systems that will produce an electromagnetic counterpart for different values of $R_\mathrm{NS}$. All the rates shown on the upper panels of Fig.~\ref{fig:rates} are also summarized in  Table~\ref{table:rates}. The most striking feature of Fig.~\ref{fig:rates} is that detection rates from the N20 engine (the purple diamonds) are higher by a factor of $\sim 2-10$  with respect to the predictions for the ``rapid'' and ``delayed'' mechanisms in all models. This is because the N20 engine predicts the formation of low mass BHs by direct collapse. These BHs  do not receive natal kicks, that would otherwise disrupt the binary, and they are produced by less massive stars whose number is favored by the initial mass function. 

To explore further the role of BH natal kicks on the detection rates of BHNS mergers, we focus on the NO-BH-KICK model (upper right panel in the Fig.~\ref{fig:rates}). The assumption of no BH kicks increases the detection rate for the ``delayed'' mechanism by a factor of ${\sim}4$ with respect to the STANDARD model. In contrast the rate for the ``rapid'' prescription does not have a significant increase between models. This difference is due to the fact that the pre-core-collapse He-core mass, $\Mhec$, range where the ``rapid'' mechanism predicts BH formation via partial fall-back and non-zero kick velocities is smaller ($\Mhec\in[9,13] \Msun$) compared to the ``delayed'' mechanism  ($\Mhec\in[6,13] \Msun$), as shown in Fig.~\ref{fig:N20He}. Furthermore, since this region is located at higher $\Mhec$ values for the ``rapid'' mechanism, the steepness of the initial mass function results, in any case, in less BHs with non-zero kicks. Despite all of this, the N20 engine produces more than twice as many BHNS mergers, even when only considering the NO-BH-KICK models. This is explained by the fact that low mass BHs formed by the ``delayed'' mechanism in the NO-BH-KICK model are not produced by direct collapse. Meaning that only a fraction (as low as $\sim 30\%$) of the pre-SN mass of their progenitors will collapse and the orbit of the binary needs to be readjusted even if the BH does not receive a natal kick. The readjustment of the orbit in such cases reduces the number of BHNS mergers with respect the amount expected if the low mass BHS were produced by direct collapse.

\subsection{Rates of electromagnetic counterparts linked to BHNS mergers}

\begin{figure}
\centering
%\captionsetup{justification=justified}
%\justify
\includegraphics[width=0.49\textwidth]{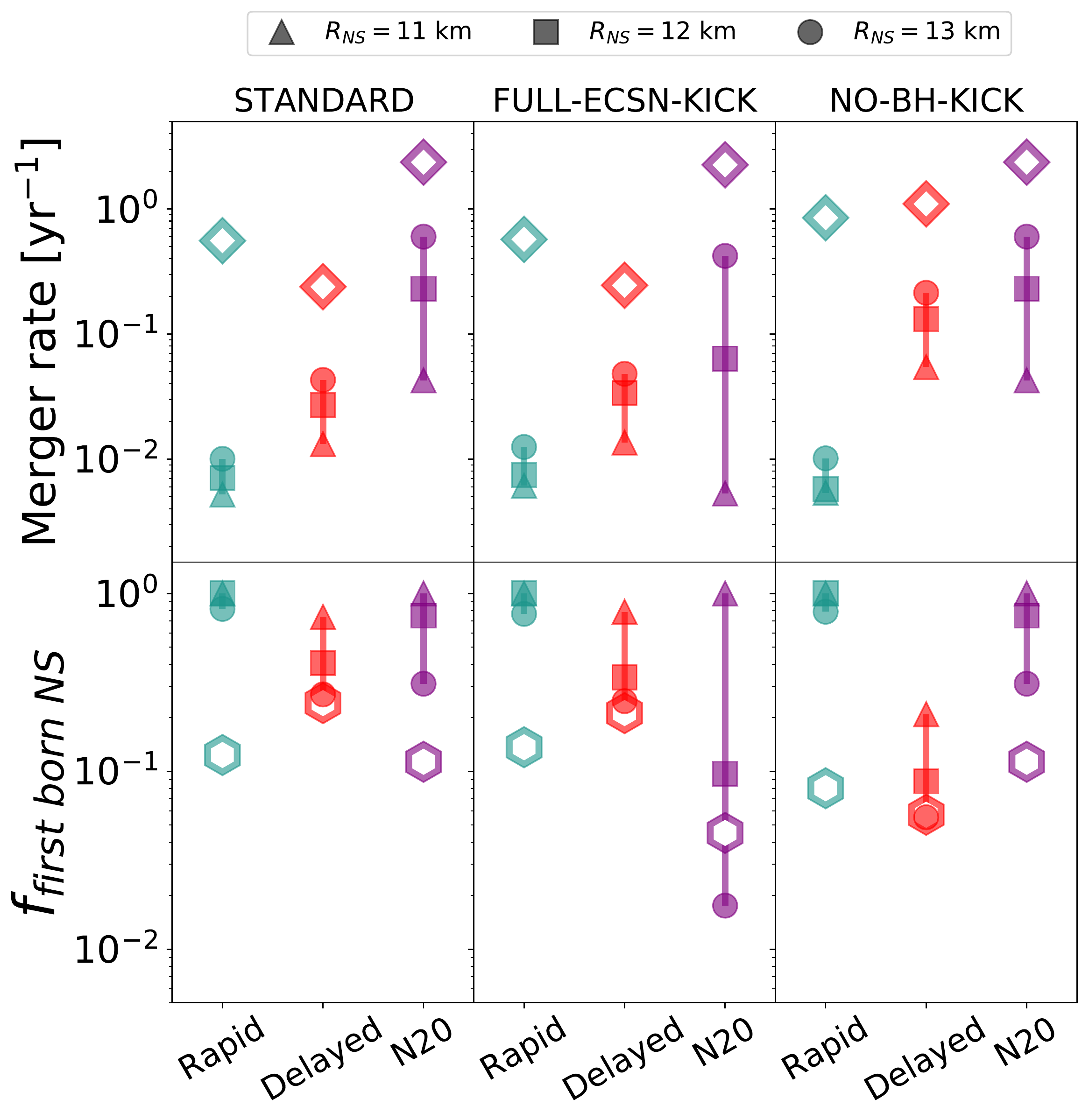}
\caption{Predicted merger rate of BHNS populations for O3a (upper panels) and the fraction of systems with first-born NSs $f_\mathrm{first\ born\ NS}$ (lower panels). This predicted rates are plotted each core-collapse mechanism mechanisms and each model respectively, see labels. In the upper panels, the unfilled diamonds indicate the predicted detection rate for the whole population of BHNSs, while the filled markers indicate the EMCs considering different NS radii, see legend above. In the lower panels, the unfilled hexagons denote the detected fraction of BHNSs with first-born NSs with respect to the whole population of detected mergers, while the filled markers are the fractions of EMC from systems with firstborn NSs with respect to all of them, see legend above for the shape description.}

 \label{fig:rates}
\end{figure}

The predicted rates of EMCs considering three different values for NS radii, 11, 12 and 13\,km, are shown in the upper panels of Fig.~\ref{fig:rates} with the triangle, square and circle markers, respectively. The predicted EMC rates for the N20 engine in the STANDARD population remains higher than the ones predicted using the other core-collapse prescriptions and, for all NS radii. Note, however, that it is the ``delayed'' mechanism that has the largest fraction of BHNS mergers with EMCs. This can be understood from the fact that among the three mechanisms considered, it is only the ``dalayed'' that can produce low-mass BHs, in the range of $2.5-3.5\rm\, M_{\odot}$. These low-mass BHs, even when they are non-spinning, are able to disrupt a relatively compact ($\lesssim 12\,\rm km$) NS outside their ISCO.

For the ``rapid'' and N20 engines and assuming NS radii $\lesssim 12\,\rm km$, BHNS mergers with EMCs are only produced when the NS is the first-born compact-object which subsequently tidally spins up the BH progenitor star and produces a higly spinning BH. The bottom row of Fig.~\ref{fig:rates} shows that although the fraction of first-born NS in BHNS merges, for the ``rapid'' and N20 engines, is $\sim 10\%$, the fraction of BHNS mergers with EMC that had a first-born NS is close to 100\%. 

The formation of the first compact object occurs in a wide orbit. Since the probability of the binary to remain bound after a SN kick scales with the ratio of the orbital velocity over the kick velocity \citep{kalogera1996orbital}, a NS formed via FeCCSN, that typically receives a natal kick of hundreds of km\,s$^{-1}$, will disrupt the binary. With the N20 engine, however, as BHs may be produced by $\Mhec$ as low as $\sim4.5\rm\, M_{\odot}$, a non-negligible fraction of first-born NSs may come from progenitors with low enough pre-core-collapse $\Mhec$ that results to an ECSN. The low kicks associated with ECSN increase significantly the survivability of these relatively rare binary configurations. 

The FULL-ECSN-KICK models allow us to quantitatively explore the role of ECSN on the formation of BHNS with EMCs, which as explained above affects models with the N20 engine. Increasing the ECSN kick velocities has primarily an impact on the systems with first-born NSs rather than the whole population of BHNSs. From Table~\ref{table:rates}, we see that the BHNS merger detection rate slightly decrease for N20 from STANDARD to FULL-ECSN-KICK. Contrarily, such rates for the other core-collapse mechanisms remain similar, as the fraction of first-born NSs from ECSN is negligible in those cases\footnote{In fact, we do see a very small increase in the detection rate for those models, which is however above the Poisson error of our simulations. This very small increase may be explained by the fact that in those models some of the second-born NSs are produced by ECSN. Imparting a larger natal kick on those NSs will not disrupt the binary as its orbit is very close at that point in time, but will impart some eccentricity in the post-SN orbit which may shorten the time to merge due to GW emission. This in turn can have a small effect on the overall rate.}. Most importantly though, high ECSN kicks decrease the rate of BHNS with EMCs, in the N20 engine by a factor of $\sim 4$, for NS with radii $\lesssim 12\,\rm km$, see Table~\ref{table:rates}.

%%%%%%%%%%%%%%%%%%%%%%%%%%%%%%%%%%%%%%%%%%%%%%
%%%%%%%%%%%%%%%%%%%%%%%%%%%%%%%%%%%%%%%%%%%%%%%%% 

Current constraints on the NS equation of state indicate that NS radii can be as high as $13$ km \citep[see][ and references therein]{capano2020stringent,chatziioannou2020neutron}. This parameter has a crucial impact on the population of EMCs from BHNS mergers where the BH is the first-born compact object. In such cases the BH spin is negligible; therefore a $R_\mathrm{NS}$ on the high end of the allowed parameter space is  crucial to achieve a NS tidal disruption \citep{foucart2018remnant}. In fact, for $R_\mathrm{NS}=13\,\rm km$ the maximum BH mass that can tidally disrupt a NS is below the minimum BH mass predicted by the N20 engine. This allows for BHNS mergers with EMCs originating from binaries with non-spinning, first-born BHs and translates to an increase by a factor of 15 and 2.5 of EMCs, compared to assumed  $R_\mathrm{NS}$ of 11\,km and 12\,km, respectively.

\begin{table*}
\caption{The predicted local detection rates, $\mathcal{R}^\mathrm{det}$ (detected mergers per year), for the LVC's third observing run. Results here are for the whole population of BHNSs (All) the electromagnetic counterparts (EMC), and for systems like GW190426\_152155 (GW190426-like), and the subset of GW190425-like systems those systems that will produce an electromagnetic counterpart (GW190426 EMC) for each physical model.   \label{table:rates}}
\centering
\begin{tabular}{ccccccccccccccc}
%\toprule

\cmidrule{4-6}
\cmidrule{8-10}
\cmidrule{12-14}
     &  & &  \multicolumn{3}{c}{STANDARD}  & &   \multicolumn{3}{c}{FULL-ECSN-KICK} & &   \multicolumn{3}{c}{NO-BH-KICK}  \\

\cmidrule{4-6}
\cmidrule{8-10}
\cmidrule{12-14}

Population     & $R_\mathrm{NS}$ &  &  $\mathcal{R}^\mathrm{det}_\mathrm{N20}$ &  $\mathcal{R}^\mathrm{det}_\mathrm{Rapid}$ &  $\mathcal{R}^\mathrm{det}_\mathrm{Delayed}$ & &  $\mathcal{R}^\mathrm{det}_\mathrm{N20}$ &  $\mathcal{R}^\mathrm{det}_\mathrm{Rapid}$ &  $\mathcal{R}^\mathrm{det}_\mathrm{Delayed}$  & &  $\mathcal{R}^\mathrm{det}_\mathrm{N20}$ &  $\mathcal{R}^\mathrm{det}_\mathrm{Rapid}$ &  $\mathcal{R}^\mathrm{det}_\mathrm{Delayed}$  \\
\toprule
%\multicolumn{5}{c}{STANDARD} \\
%\midrule
All &  &  & 2.37 &	0.56 &	0.24 & &	2.25 &	0.57 &	0.25 & &	2.37 &	0.85 &	1.10\\

 \multirow{3}{*}{EMC}& $11$km &  &   0.04 &	0.01 &	0.01& 	&0.01& 	0.01& 	0.01& 	&0.04& 	0.01& 	0.05  \\
 & $12$km &  &   0.23& 	0.01& 	0.03& 	&0.06& 	0.01& 	0.03& 	&0.23& 	0.01& 	0.13 \\
 & $13$km &  &   0.6& 	0.01& 	0.04& &	0.42& 	0.01& 	0.05& &	0.6& 	0.01& 	0.21\\
 
  %&  &  &  & & & & & & & & & & \\
\midrule
  
  GW190426-like &  &  & 0.56 &	0.03 &	0.06 & &	0.55 &	0.05 &	0.07 & &	0.56 &	0.12 &	0.42\\
  
   \multirow{3}{*}{GW190426 EMC}& $11$km &  &   0.00 & 0.00 & 0.00 & &   0.00 & 0.00 & 0.00 & &   0.00 & 0.00 & 0.00\\
   & $12$km &  &   0.01 & 0.00 & 0.00 & &   0.00 & 0.00 & 0.00 & &   0.01 & 0.00 & 0.00\\
 & $13$km &  &   0.05& 	0.00 &	0.02& &	0.03 &	0.00 &	0.01 &	&0.05 &	0.00 &	0.03 \\
\bottomrule
\end{tabular}
\end{table*}

\subsection{The role of the SN engine on the synthesis of GW190426\_152155-like events}

The estimation for the BHNS merger rate form the LVC O3a run, $2$ -- $4$ yr$^{-1}$, is closer to the results from the N20 engine on all populations, as well as for the ``delayed'' for the NO-BH-KICK population, favoring the engines that predict low-mass BHs with negligible SN kicks. We compute the merger rate of systems like GW190426\_152155, the only observed system labeled as a BHNS merger \citep{Abbott:2020niy}, by considering the $90\%$ confidence intervals of the event's measured total mass, chirp mass and the effective spin. Those results are shown in Table~\ref{table:rates} as the population ``GW190426-like''. The detection of a system similar to GW190426\_152155 is favored by the N20 engine in all populations and by the ``delayed'' mechanism in the NO-BH-KICK populations; while the ``rapid'' prescription predicts a rate of only $0.03$ -- $0.12$ yr$^{-1}$, indicating that the detection of an event with GW190426\_152155-like properties is rare in such case.

We also computed the rate of GW190426\_152155-like systems that will produce an EMC (population ``GW190426 EMC'' in Table~\ref{table:rates}). In all cases the rates of EMCs linked to events like GW190426\_152155 are less than $0.1$ yr$^{-1}$, translating to a probability of $0$ -- $25$ \% of EMCs per observed system. This low rate agrees with the lack of an electromagnetic signal linked to the observed event.

\subsection{Comparison to BBH merger density rates}

In order to compare our models with measurements with better constraints, we compute the local BBH and BHNS merger density rates as shown in Table~\ref{table:dens} of Appendix~\ref{app:dens}.
The N20 STANDARD model, which increases the likelihood of events like GW190426\_152155, over-predicts the measured BBH merger density, as reported by \cite{Abbott:2020niy}, by $\sim 5$ times compared to the other collapse mechanisms which over-predict the rate by $< 2$ times.
Again, this is because the N20 engine assumes BH are born without kicks. Similarly, ``rapid'' and ``delayed'' NO-BH-KICK over-predict the local BBH rate by the same amount. When looking carefully at the BBH's progenitor evolutionary pathway we find that it is the CE channel that produces the majority of these merging BBHs contributing to the observed local rate density. If, indeed, BHs are born without a kick then the models are overproducing the systems going trough and surviving CE and therefore the population of BBH, as we expect the mixture of all formation channels to contribute on the whole population, which is what recent studies suggests \citep[e.g.][]{2016A&A...596A..58K, 2017MNRAS.465.2092P, 2020arXiv200611286K}.

\section{Discussion and conclusions \label{sec:conclusions}}
In this work, we present population synthesis models exploring the role of core-collapse prescriptions, including results of detailed 1D core-collapse simulations, and the associated natal kicks, on the observability of BHNS systems, events like GW190426\_152155, and their EMCs. A parametric code has been used to model binary stars from ZAMS until the first-born compact object strips its companion star. From that stage, the evolution of the system until the formation of the second-born compact object was followed by detailed binary-evolution models, allowing for accurate predictions of the spin of the second-born compact object. The latter is critical in determining whether the NS will be tidally disrupted by the BH, as spinning BHs increase significantly the probability of EMCs produced by BHNS mergers.

We find that the N20 SN engine predicts BHNS merger rates higher by an order of magnitude compare to the ``rapid'' and ``delayed'' mechanisms. This is a consequence of the formation of low-mass BHs by direct collapse that are predicted by SN engines from detailed core-collapse simulations, such as N20. In addition, the N20 engine predicts higher rates of EMCs while being consistent with the lack of observations of such events to date. 

Our models show that future, more stringent constraints on the NS equation of state will allow to distinguish between formation sub-channels of BHNSs. A mean NS radius closer to $11$ km would indicate that the information from future observations of BHNS EMCs is linked to systems where the NS is the first-born compact object; as compact NSs are harder to disrupt by non-spinning BHs. In contrast, evidence of a larger radii \citep[such as][]{abbott2020gw190425,riley2019nicer} will help us to infer information of low-mass BHs and possible natal kicks linked to their formation.

Finally, we find that the synthesis of events like GW190426\_152155 is favored by SN engines that produce low-mass BHs with small or no SN kicks. The N20 engine predicts a rate of $\sim 0.6$\,yr$^{-1}$ for such event, one order of magnitude higher than the predictions by the ``rapid'' and ``delayed'' mechanisms in the STANDARD model, and 30\% larger than the result from ``delayed'' with the NO-BH-KICK model.  However, such mechanisms overestimate the BBH local merger density rate, with most of the predicted merging BBHs going through the CE evolution channel. Detailed binary evolution calculations suggest that the source of this apparent discrepancy may be the parametrizations of mass-transfer stability criteria and envelope binding energy estimates, which, as implemented in most binary population synthesis codes may, severely  over-predict the  number of BBH progenitor systems going through and  surviving the CE phase.

\begin{acknowledgements}

The authors thank Vicky Kalogera and Christopher Berry for their thoughtful comments. This work was supported by the Swiss National Science Foundation Professorship grant (project number PP00P2 176868; PI Tassos Fragos). JRG is supported by UNIGE, JJA and SC are supported by CIERA and AD, JGSP, and KAR are supported by the Gordon and Betty Moore Foundation through grant GBMF8477. KK received funding from the \it{ European Research Council} under the European Union's {\it Seventh Framework Programme} (FP/2007-2013) / {\it ERC} Grant Agreement n.~617001. YQ acknowledges funding from the Swiss National Science Foundation under grant P2GEP2\_188242. The computations were performed in part at the University of Geneva on the Baobab and Lesta computer clusters and at Northwestern University on the Trident computer cluster (the latter funded by the GBMF8477 grant). All figures were made with the free Python modules Matplotlib \citep{hunter2007matplotlib}. This research made use of Astropy,\footnote{http://www.astropy.org} a community-developed core Python package for Astronomy \citep{robitaille2013astropy,price2018astropy}.

\end{acknowledgements}

%\bibliographystyle{aa} % style aa.bst
\bibliographystyle{aasjournal}
\bibliography{ref} % your references Yourfile.bib
\appendix
\section{Implementation of the N20 engine}
\label{app:N20}

We consider the stellar models by \cite{sukhbold2016core}, evolved in solar metallicity, to describe the baryonic mass of the final remnant. We apply the prescription deduced from their results to stars with different metallicity as we aim to explore the effect of the average non-monotonic explodability trends rather than predicting accurately the compactness of the core. Such non-monotonic trend is preserved for different metallicities \citep{patton2020towards}.

\begin{figure}
\centering
%\captionsetup{justification=justified}
%\justify
\includegraphics[width=0.49\textwidth]{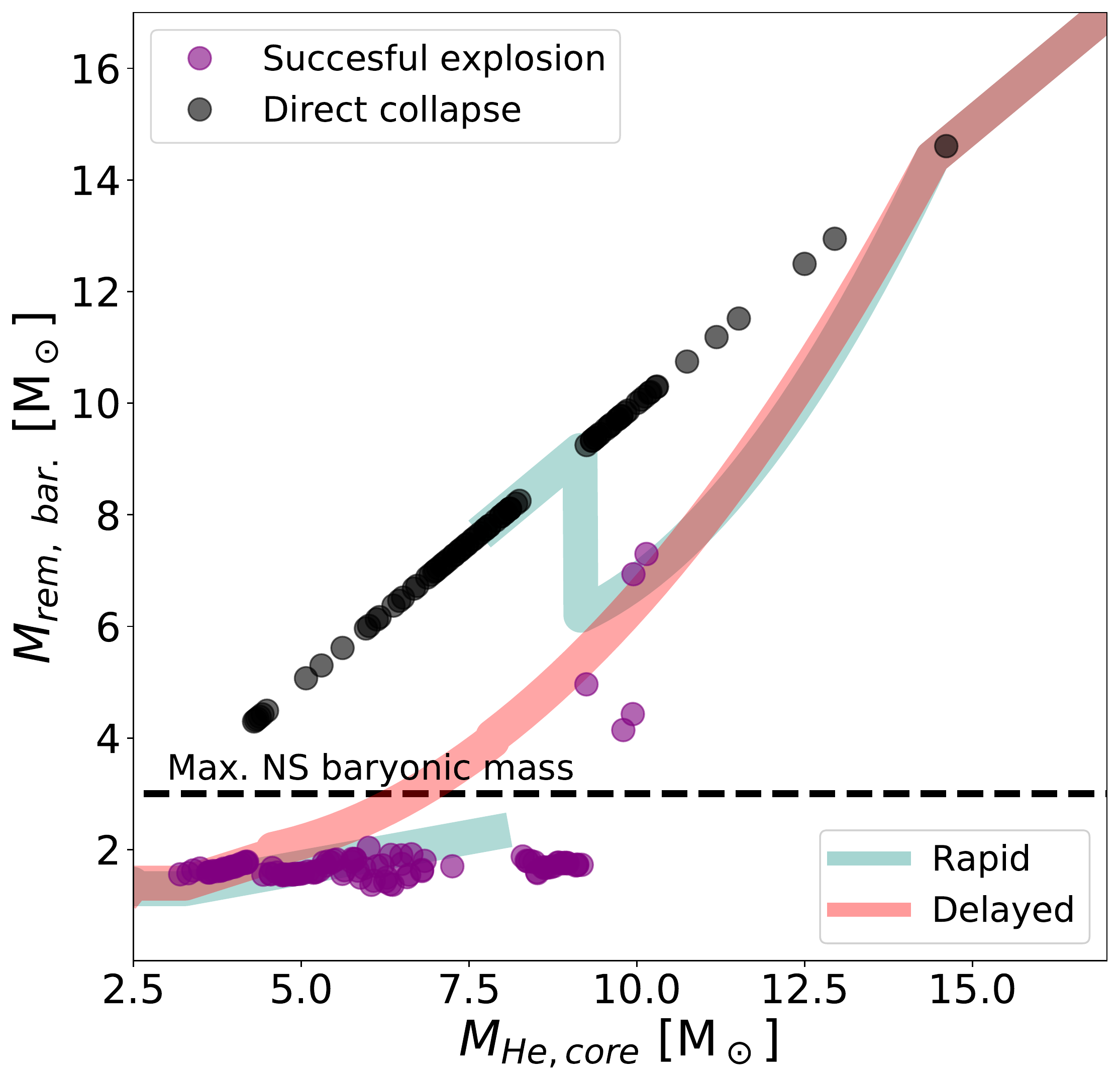}
\caption{Baryonic remnant mass, $M_\mathrm{rem,\,bar}$, as a function of the He-core mass at the pre-supernova phase, $\Mhec$, from N20 engine of \cite{sukhbold2016core} (circles); and for the ``rapid'' and ``delayed'' (red and turquoise thick lines) mechanisms from \cite{fryer2012compact}. For illustrative purposes we assume that the mass of the carbon-oxygen core is $0.76 \Mhec$.  Each model exploded with the N20 engine is labeled as a successful {explosion} or direct collapse (i.e. failed explosion), with purple and black colors, respectively. The black dashed line represents the maximum NS baryonic mass, in our model, which corresponds to a maximum NS gravitational mass of $2.5\,\Msun$.}
 \label{fig:N20He}
\end{figure}

Fig.~\ref{fig:N20He} shows the baryonic remnant mass as a function of $\Mhec$, for the N20 engine \citep[][]{sukhbold2016core}, the ``rapid'' and ``delayed'' mechanisms \citep{fryer2012compact}. On the same figure, each stellar model, collapsed with N20, is classified as a successful explosion or a direct collapse depending if the supernova shock is reactivated by the neutrino flux or not, respectively. In the case of the N20 engine, to predict if a star will undergo a successful explosion or direct collapse, we extract the result in terms of the nearest-neighbor of the star's $\Mhec$ with respect to the results from \cite{sukhbold2016core}. If the star is classified as a progenitor of a successful explosion, its $M_\mathrm{rem,\,bar}$ will have the same value of the remnant baryonic mass associated to the point with the nearest value for $\Mhec$. Otherwise, if the star is classified as a progenitor of a direct collapse, then $M_\mathrm{rem,\,bar}$ is equal to the pre-SN mass of the star.

To determine if the remnant is a BH or a NS, we calculate the remnant gravitational mass considering the neutrino loss as in \cite{zevin2020exploring}, where, here, the maximum mass loss by neutrinos is considered to be $0.5\,\Msun$. If the remnant gravitational mass is larger than $2.5\, \Msun$ we assume that the compact object is a BH, else a NS. For this work, the successful explosions that produce massive remnants that will end up as BHs were not considered, as such cases are rare \citep[from][only 5 models from 105 successful explosions form a BH for the N20 engine]{sukhbold2016core}.

%%%%%%%%%%%%%%%%%%%%%%%%%%%%%%%%%%%%%%%%%%%%%%
%%%%%%%%%%%%%%%%%%%%%%%%%%%%%%%%%%%%%%%%%%%%%%%%% 

\section{Grids of detailed NS/BH--He-star models}
\label{app:grid}

\begin{figure*}
\centering
\includegraphics[width=18cm]{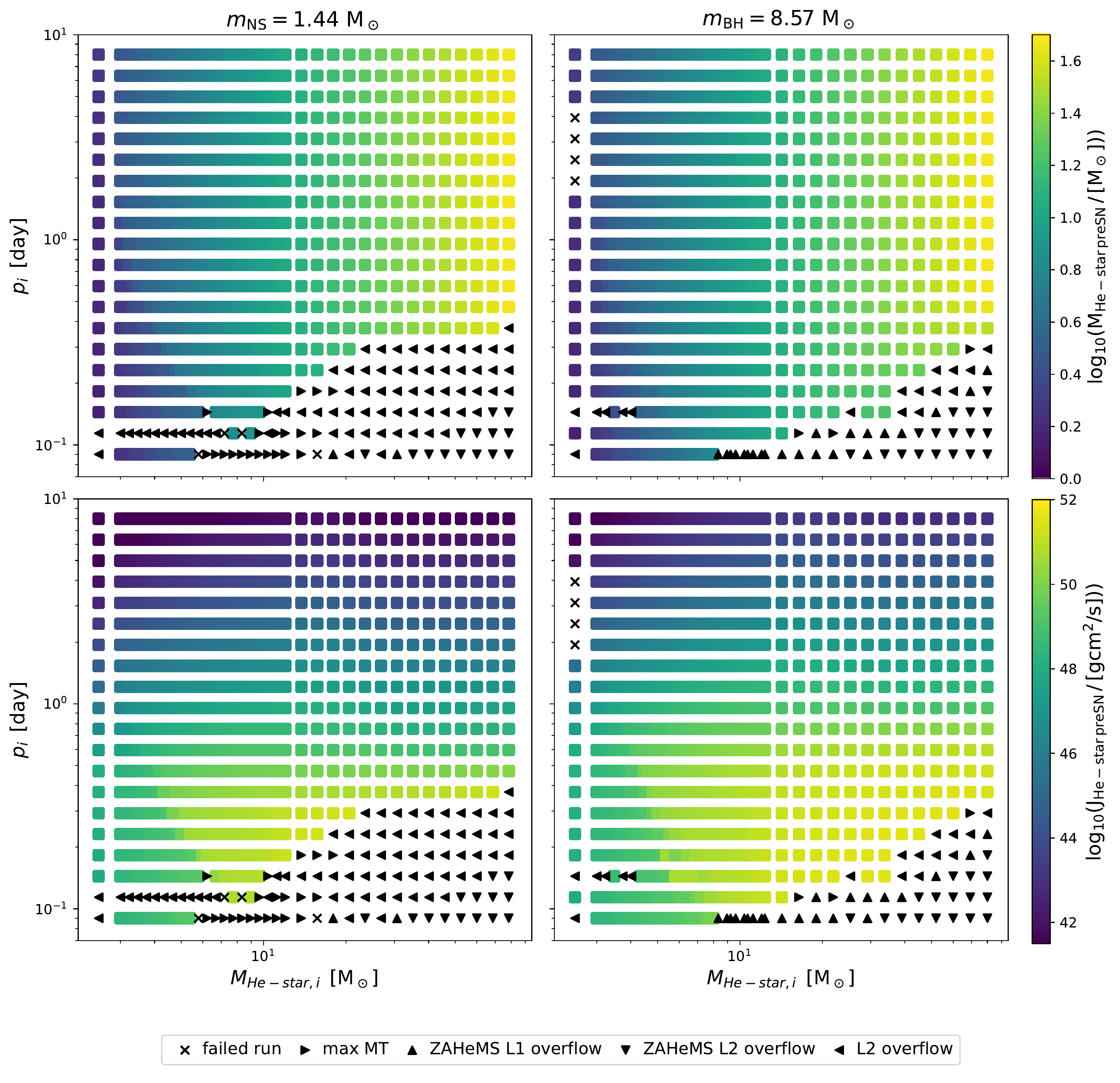}
 \caption{Two 2D slices of the 4D \mesa{} grid for $Z=0.00312$, and  $m_\mathrm{NS}=1.44\,\mathrm{M}_\odot$ and $m_\mathrm{BH}=8.57\,\mathrm{M}_\odot$, respectively. The final He-star mass and angular momentum pre-SN values are colored for each successful track according to each color bar. Each successful run stopped because of carbon depletion or off-centre neon ignition (square markers), while other termination flags are shown in the bottom legend. For visualisation purposes, the models at $p=0.04$ days were excluded from the figure which for these slices all runs resulted in ZAHeMS L1 or L2 Roche-lobe overflow.
 }
 \label{fig:MESA}
\end{figure*}

\begin{figure}
\centering
\includegraphics[width=\columnwidth]{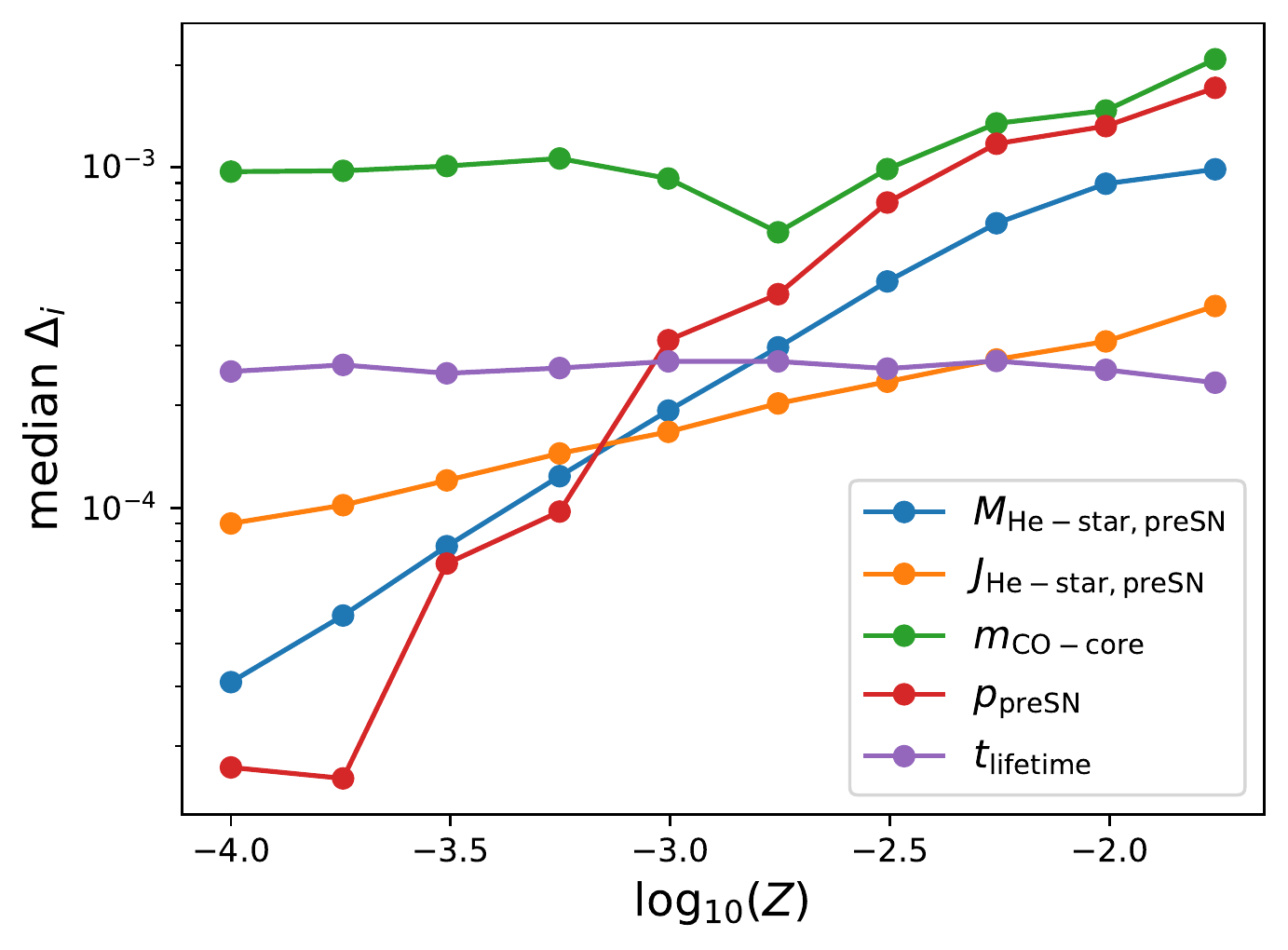}
 \caption{Median relative error as a function of metallicity of the log-transformed and re-scaled five quantities $A_i$: the He-star mass (blue), angular momentum (orange) and its carbon-oxygen core mass (green) before the supernova, the orbital period before the supernova (red) and the lifetime of the binary (green).
 }
 \label{fig:grids-errors}
\end{figure}

We use subset of detailed \mesa{} models of BH--He-star systems of \cite{2020arXiv201016333B} which we extend to cover the NS star and low He-star masses. These models treat the compact object as a point mass, hence, they can be applied to simulate NS--He-star systems. For simplicity, we assumed that all physical assumptions made in the BH--He-star regime applies also to the NS--He-star regime too, including an Eddington mass-accretion rate limit of $\dot{M}_\mathrm{Edd} = 7.36 \times 10^{-8} \left(M_\mathrm{BH/NS} / \mathrm{M}_{\odot}\right)\,\mathrm{M}_\odot{}\mathrm{yr}^{-1}$. 
The selected subset of the original grid covers the initial parameter space of 10 metallicities, $Z$, in the log-range $[0.0001,0.0174]$ in log-steps of $\Delta \log_{10}(Z) \simeq 0.25$, 11 BH masses in the log-range $[2.5, 54.4] \, \mathrm{M}_\odot$, 17 He-star masses in the log-range $[8, 80] \, \mathrm{M}_\odot$ and $20$ binary periods in the log-range $[0.09, 8] \, \mathrm{days}$. We extend this dataset to cover 10 NS masses in the log-range $[1,2.28]\,\mathrm{M}_\odot$, 26 He-star masses where 20 are in the log-range $[3,12]\,\mathrm{M}_\odot$, 5 in the range $[3,7]\,\mathrm{M}_\odot$ (only for BH masses) and one at $2.5\,\mathrm{M}_\odot$ and an extra period at 0.04 days. The smallest He-star mass is chosen to guarantee coverage of the parameter space down to white dwarf formation, while the maximum He-star mass and smallest orbital period were chosen to include the full range of compact object -- He-star systems produced by our \cosmic{} models. The wide orbital period range ensures that we cover the parameter space well past the point where any BHNS or BBH system will merge within the age of the Universe. The original grid subset consisting of 37,400 models was therefore extended to a total of 172,740 \mesa{} models. The fraction of failed \mesa{} runs vary from $0.6\%$ to $1.5\%$ depending on metallicity. In Figure~\ref{fig:MESA} we show two 2D slices of the 4D parameter space sliced at $Z=0.00312$ and, $m_\mathrm{NS}=1.44\,\mathrm{M}_\odot$ and $m_\mathrm{BH}=8.57\,\mathrm{M}_\odot$, respectively, where we indicate with a color the final He-star mass and angular momentum given initial orbital separation and He-star masses of the detailed simulations. 

These grids were used to determine the final outcomes and final  parameters of the late-end evolution stage of the binary systems through linear interpolation. Each metallicity is interpolated separately. We want to interpolate five quantities $A_i$:  the He-star mass, angular momentum and its carbon-oxygen core mass, orbital period before the supernova, as well as the lifetime of the binary. Before interpolating each quantity, we log-transformed it and re-scale it to the interval $[-1,1]$ to assign equal weight to each dimension during the interpolation. The interpolation itself relies on building a Delaunay triangulation of the input data points followed by barycentric linear interpolation over the vertices of the (hyper)triangle containing the location of interest. We test the accuracy of the interpolation computing relative errors of a test grid which is constituted of an arbitrary fraction (5\%) of runs which we excluded from the train sample. To obtain a consistent estimate, we repeat this experiment 10 times for each metallicity and each interpolated quantity. If a point of an nonphysical region of the parameter space (e.g. zero age He main sequence (ZAHeMS) overflow, max MT or L2 overflow) is correctly interpolated to NaN (not a number) by the algorithm, we consider it to have a zero relative error. On the other hand, if a point is wrongly interpolated to NaN, we consider it to have a relative error of 1. In Fig.~\ref{fig:grids-errors} we report the median relative error of each transformed and rescaled quantity $X_i\equiv \log_{10}(A_i)^{[-1,1]}$ as $\Delta_i = |X_\mathrm{true,i}-X_\mathrm{interp,i}|/X_\mathrm{true,i}$. Because of the large sample of data-points we find small interpolation errors. Most of the quantities show an increase of median relative error as a function of metallicity. This is caused by the fact that at high metallicity the grids show a less-linear behaviour than at low metallicity. This non-linearity is a direct consequence of He-star stellar winds which, in our models, scale as $ (Z/Z_\odot)^{0.85}$. In these systems, the He-stars lose a non-negligible amount of mass and the orbits widen considerably. Moreover, NS--He-star in tight orbits have higher probability to initiate a mass transfer case due to the He-star tendency to expand more than at low metallicity.

\section{BHNS electromagnetic counterpart condition}
\label{app:EMC}

We consider the model of \citet{foucart2018remnant} to determine the mass of the NS that remains outside the BH ISCO after the tidal disruption, as

\begin{equation}
    \left[\max \left(\alpha \frac{1-2C_\mathrm{NS}}{\eta^{1/3}}{-}\beta \hat{R}_\mathrm{ISCO} \frac{C_\mathrm{NS}}{\eta}{+}\gamma, 0\right)\right]^{\delta} M_\mathrm{NS}
\end{equation}
where $\alpha{=}0.406$, $\beta{=}0.139$, $\gamma{=}0.255$, $\delta{=}1.761$, $\hat{R}_\mathrm{ISCO}{\equiv}\frac{R_\mathrm{ISCO} c^2}{G M_\mathrm{BH}}$, $C_\mathrm{NS}{=}Q \frac{R_\mathrm{ISCO}}{\hat{R}_\mathrm{ISCO} {R}_\mathrm{NS}}$, $Q{=}\frac{M_\mathrm{BH}}{M_\mathrm{NS}}$, $\eta = Q/(1+Q)^2$. Here, $M_\mathrm{BH}$ is the mass of the BH, $M_\mathrm{NS}$ the mass of the NS, and $R_\mathrm{NS}$ is the radius of the NS. In this work we explore three values for $R_\mathrm{NS}$, as 11, 12, 13, km.

\label{app:dens}
\begin{table*}
\caption{ Predicted local rate density $\mathcal{R}$ (in units of $\mathrm{Gpc}^{-1}{}\mathrm{yr}^{-1}$). Results here are shown for the common envelope (CE) and stable-mass transfer (SMT) channels sepratatly and combined for, both, the populations of BBH and BHNS for each model.  \label{table:dens}}
\centering
\begin{tabular}{ccccccccccccccc}
%\toprule

\cmidrule{4-6}
\cmidrule{8-10}
\cmidrule{12-14}
     &  & &  \multicolumn{3}{c}{STANDARD}  & &   \multicolumn{3}{c}{FULL-ECSN-KICK} & &   \multicolumn{3}{c}{NO-BH-KICK}  \\

\cmidrule{4-6}
\cmidrule{8-10}
\cmidrule{12-14}

    & Channel &  &  $\mathcal{R}_\mathrm{N20}$ &  $\mathcal{R}_\mathrm{Rapid}$ &  $\mathcal{R}_\mathrm{Delayed}$ & &  $\mathcal{R}_\mathrm{N20}$ &  $\mathcal{R}_\mathrm{Rapid}$ &  $\mathcal{R}_\mathrm{Delayed}$  & &  $\mathcal{R}_\mathrm{N20}$ &  $\mathcal{R}_\mathrm{Rapid}$ &  $\mathcal{R}_\mathrm{Delayed}$ \\
\toprule
%\multicolumn{5}{c}{STANDARD} \\
%\midrule

 \multirow{3}{*}{BBH}& CE &  &   162.78 &	81.13 &	39.13 & &	162.06 & 81.64 &	38.55 &	& 162.78 &	129.41 &	167.31\\
 & SMT &  &   39.73	& 33.05	& 31.27& &	41.43&	33.6&	31.08& &	39.73&	31.38&	26.36 \\
 & CE + SMT &  &  202.51 &	114.18	& 70.4	& &203.49&	115.24&	69.63& &	202.51&	160.79&	193.67 \\
 
  %&  &  &  & & & & & & & & & & \\
\midrule

   \multirow{3}{*}{BHNS}& CE &  &   74.96&	7.28	&2.59& &	66.36&	7.59&	2.93	& &74.96&	14.66&	30.18\\
   & SMT &  &   2.28&	2.86&	3.06& &	2.99&	3.12&	2.88& &	2.28&	1.88&	2.49\\
 & CE + SMT &  &  77.24 &	10.14&	5.65& &	69.35&	10.71&	5.81& &	77.24&	16.54&	32.67 \\
\bottomrule
\end{tabular}
\end{table*}

\section{BBH and BHNS merger rate densities}

To better understand the implication of the model physical modelling assumptions done in this study, we also calculate the BBH merger rate densities.  The latter are much better constrained by GWTC-2, which finds the local merger rate density of BBH to be $23.9^{+14.9}_{-8.6} \, \mathrm{Gpc}^{-3} \mathrm{yr}^{-1}$ \citep{2020arXiv201014533T}. The BBH rate densities rising from the common envelope (CE) and stable mass transfer (SMT) channels given the same set of assumption made here (STANDARD-Delayed) are presented in \cite{2020arXiv201016333B}. For an one-to-one comparison with BHNS rate densities of the different core-collapse and kick prescriptions considered in this work, we summarise the rate densities of BBHs and BHNS for the CE and SMT channels as well as their combination in Table~\ref{table:dens}. 

In contrast to the original study done by \cite{2020arXiv201016333B} the new models differ in the following ways. First, we simulated a (i) metallicity range with one third the resolution of \cite{2020arXiv201016333B} but verified that this does not have a noticeable impact on the rate estimates by reanalysing \cite{2020arXiv201016333B} models with the same metallicity sample resolution. Second, (ii) the core-collapse of the secondary is assumed to be direct where in the original study we followed the core-collapse of the \mesa{} He-star profile at SN accounting for disk formation. When an accretion disk is formed only a fraction of its mass falls to the hole \citep[see Appendix~D in][]{2020arXiv201016333B} which, in practice, means that here tidally spin up highly spinning BHs are slightly more massive compared to \cite{2020arXiv201016333B}. Moreover, we only (iii) interpolate binary properties before the supernova while in the original work, which only investigated the delayed collapse mechanism, also interpolated the second born compact object mass and spin (the former has, on average, a larger interpolation error compared to the pre SN mass, see Fig.~E.1 of \citet{2020arXiv201016333B}). Finally, we also updated the condition which determines BH formation. We assume (iv) a BH is formed if the compact object gravitational mass is larger than the maximum NS mass ($2.5 \, \mathrm{M}_\odot$) while in the previous work we assumed that a collapsing star leading to the formation of a BH had to have at least a carbon-oxygen core mass and a remnant baryonic mass of $3\,\mathrm{M}_\odot$ in order to form a BH (as $0.5 \, \mathrm{M}_\odot$ where assumed to be lost because of neutrinos during the collapse of the proto-NS). We verified that this change has no impact on the results.

In Table~\ref{table:dens} we can see that the STANDARD-delayed model predicts a comparable rate density $\sim{}39\,\mathrm{Gpc}^{-3}\mathrm{yr}^{-1}$ and $\sim{}31\,\mathrm{Gpc}^{-3}\mathrm{yr}^{-1}$ for CE and SMT channels, respectively. These values are in agreement with \cite{2020arXiv201016333B} where the small deviation in the numbers is given by the changes explained in the previous paragraph. On the other hand, STANDARD-N20 overpredicts the CE+SMT rates compared to the observations by a factor of at least 5. This is because in the N20 engine all BHs are formed thorough direct collapse, without a kick. When assuming no natal kicks (other than the readjustment of the orbits because of neutrino mass loss) the NO-BH-KICK models with rapid and delayed predict similar rates to the N20 engine meaning that the discrepancy is a direct product of no BH kicks. In fact, if nature would to agree with the N20 engine, then the formation of merging BBH thorough the CE+SMT over-predicts the systems surviving these channels. When looking more carefully at the rate densities of these models we see that is the CE channel to over-predicting the constrains from merging BBH in the local universe. As recent studies have shown, the classical $\alpha_\mathrm{CE}-\lambda$ parameterization of CE and mass transfer stability parameterization ($q_\mathrm{crit}$) might over-predict the number of systems going thorugh and surviving this evolutionary phase \citep[e.g.][]{2016A&A...596A..58K, 2017MNRAS.465.2092P, 2020arXiv200611286K}.

%%%%%%%%%%%%%%%%%%%%%%%%%%%%%%%%%%%%%%%%%%%%%%%%%%%%%%%%%%%%%%%%%%%%%%%%%%%%%%%%%%%%%%%%%%%%

%\bibliography{ref}
%\bibliographystyle{aasjournal}



\end{document}